\documentstyle[12pt]{article}
\begin{document}
\titlepage
\title{
\begin{flushright}
Preprint PNPI-2010, November 1994.
\end{flushright}
\bigskip\bigskip\bigskip
2d gravity stress  tensor and the problem of the
calculation of the multi-loop amplitudes in the string
theory}
\date{}
\author{G.S. Danilov\thanks{E-mail address: danilov@lnpi.spb.su}\\
Petersburg Nuclear Physics Institute,\\
Gatchina, 188350, St.-Petersburg, Russia}
\maketitle
\begin{abstract}
It is shown that the vacuum value of the $2d$ gravity stress
tensor in the free field theory is singular in the fundamental
region on the complex plane where the  genus-$n>1$ Riemann
surface are mapped. Because of the above singularity, one can
not construct  modular invariant multi-loop amplitudes.  The
discussed singularity is due to the singularity in the vacuum
value of the $2d$ gravity field that turns out to be on the
genus-$n>1$ Riemann surfaces.
\end{abstract}

\newpage
\section{Introduction}

It is well known that the conform anomaly \cite{ggr,gsw} hampers
the construction of the string theory in  non-critical dimensions.
Because of the above anomaly, the string amplitudes in such a
theory should depend on a choice of 2-dim. fixed reference metrics.
To remove this shortcoming, it has been proposed \cite{p} to write
down the string amplitudes in the form of the
integrals over both the string fields and the 2-dim. metrics.
As far as the integration over the above metrics has been
performed, in this case the string amplitudes can not depend on
fixed reference metrics. So, the discussed theory is expected to
be invariant under both the general two-dim. reparameterization
group and the Weyl group, like the critical string one.

In the discussed scheme, it is presumed that the conform anomaly
is canceled owing to the $2d$ gravity field \cite{kpz,df,dk}.
On the classical level the action of this field has the
Liouville form. Quantum corrections do not change this form,
only the parameters in the above action are renormalized
\cite{df,dk,mm,gtw}.

The above scheme is mainly reduced \cite{kpz,df,dk,mm,gtw} to
the  free field theory. This
theory contains the string and ghost fields and the $\Phi$ field
of the $2d$ gravity, as well.
The stress energy-momentum tensor of the $\Phi$
field includes the linear in $\Phi$ term
to cancel the conform anomaly in the whole stress
energy-momentum tensor.  Owing to this term, the $\Phi$ field
receives an additional term under two-dim.  coordinate
transformations except only linear transformations.  So on the
closed Riemann surfaces of a genus-$n>1$ the discussed $\Phi$
field is necessarily receives an additional term under twists
about B-cycles. One concludes by this that the vacuum expectation
value $<\Phi>$ of the $\Phi$ field is unequal to zero.

In this paper we calculate $<\Phi>$ on the closed
Riemann surfaces of a genus-$n\geq1$ using the Schottky group
parameterization \cite{l,fms}. As in \cite{df,dk},
the conformal gauge is employed. We show that $<\Phi>$ appears
to be singular in the fundamental region $\Omega$ on the complex
plane where the Riemann surfaces are mapped.

Moreover, for $n\geq2$ the above singularity in $<\Phi>$
originates  the vacuum
expectation value of the stress tensor to be singular in the above
$\Omega$ region.  Due to the discussed singularity in the
stress tensor it appears  impossible to construct
the multi-loop partition functions, which would be consistent with
the requirement of the modular invariance.

Only the one-loop partition function can be constructed
because  in this case  the stress tensor
appears to be non-singular in the above $\Omega$ region on the
complex plane where the genus-1 Riemann surfaces are mapped.
Moreover,  one can use the linear kleinian groups
instead of the Schottky  ones by mapping the genus-1 Riemann
surfaces into parallelograms.  In this case  the  $<\Phi>$  value is
equal to zero.  The one-loop partition function in the discussed
parameterization has been calculated in \cite{gtw}. We show that the
same result is obtained also in the Schottky parameterization, the
contribution due to $<\Phi'>$ being taken into account.

The paper is organized as it follows. In Sec.2 we calculate
 $<\Phi>$, the Schottky groups being employed. We show that
 $<\Phi>$ is singular in the fundamental region on the complex
plane where the Riemann surfaces are mapped. We show also that the
above singularity for the Riemann surfaces of a
genus-$n\geq2$ originates the singularity in the vacuum value
of the stress energy-momentum tensor. In Sec.3 we discuss
the problem of the construction of the multi-loop partition
functions. The one-loop partition function is discussed in the Sec.4.

\section{Vacuum value of the $2d$ gravity stress tensor}

As it has been noted in the Sec.1, we employ the Schottky
parameterization. In this case the genus-$n$ kleinian group is
generated by the base Schottky transformations $z\rightarrow g_s(z)$
where
\begin{equation}
g_s(z)=\frac{a_s z+b_s}{c_s
z+d_3}\,,\quad a_sd_s-b_sc_s=1 \quad{\rm and}\quad s=1,2,\cdots,n\,.
\label{sch}
\end{equation}
Every  transformation (\ref{sch}) maps the
$C_s^{(+)}$ circle into the $C_s^{(-)}$ one where
\begin{equation}
C_s^{(+)}=\{z:|-c_sz+a_s|=1\}\quad {\rm and}\quad
C_s^{(-)}=\{z:|c_sz+d_s|=1\}\,.
\label{circ}
\end{equation}
So the fundamental region can be chosen to be the exterior
of all the  $C_s^{(-)}$ and $C_s^{(+)}$ circles \cite{l,fms}.
Also, it is worth-while to note that (\ref{sch}) can be rewritten
down as
\begin{equation}
\frac{g_s(z)-u_s}{g_s(z)-v_s}=k_s\frac{z-u_s}{z-v_s}
\label{mult}
\end{equation}
where $u_s$ and $v_s$ are the fixed point coordinates and $k_s$ is the
multiplier. One can think that $|k_s|<1$.

As in \cite{df,dk},  the conformal gauge is used.
The stress energy-momentum
tensor $T$ is given by
\begin{equation}
T=T_m+T_{gh}+T_\Phi
\label{tens}
\end{equation}
where $T_m$ denotes the contribution due to the
string fields, $T_{gh}$ is the ghost contribution and $T_\Phi$
is the contribution of the $2d$ gravity field $\Phi$. The $T_m$
value is given by
\begin{equation}
T_m=-\frac{1}{2}\partial X_M\partial X^M
\label{tm}
\end{equation}
where $X^M$ are the string fields, $M=1,2\cdots d$. The above
fields are normalized as ( the $M$ index is omitted )
\begin{equation}
X(z_1,\bar z_1)X(z_2,\bar z_2)\rightarrow -\ln{|z_1-z_2|^2}
\label{normx}
\end{equation}
at $z_1\rightarrow z_2$. The $T_{gh}$ is given by
\begin{equation}
T_{gh}=2b\partial c+(\partial b)c
\label{tgh}
\end{equation}
where $b$ is the tensor ghost field and $c$ is the vector ghost
one, they being normalized as
\begin{equation}
c(z_1,\bar z_1)b(z_2,\bar z_2)\rightarrow -\frac{1}{z_1-z_2}
\label{normgh}
\end{equation}
at $z_1\rightarrow z_2$. And $T_\Phi$ in (1) is equal to
\cite{df,dk}
\begin{equation}
T_\Phi=-\frac{1}{2}\partial\Phi\partial\Phi+f\partial^2\Phi\,.
\label{tphi}
\end{equation}
The $\Phi$ field in (\ref{tphi}) is normalized by the condition
that
\begin{equation}
\Phi(z_1,\bar z_1)\Phi(z_2,\bar z_2)
\rightarrow-\ln{|z_1-z_2|^2}
\label{normph}
\end{equation}
at $z_1\rightarrow z_2$.  Being
determined from the condition that (\ref{tens}) has not the conform
anomaly \cite{df,dk}, the $f$ factor appears to be
\begin{equation}
f^2=\frac{25-d}{12}\,.
\label{f}
\end{equation}
It is follows from (\ref{tphi})  that under the conformal
$z=z(\tilde{z})$ transformations the $\Phi$ field is changed as
\begin{equation}
\Phi(z)=\tilde{\Phi}(\tilde{z})-f\ln{\left|\frac{\partial z}
{\partial\tilde{z}}\right|^2}
\label{phichan}
\end{equation}
where $\tilde{\Phi}(\tilde{z})$ is the $2d$ gravity field in the
new coordinate system. So under
the kleinian group transformations (\ref{sch}) the vacuum value
$<\Phi(z,\bar z)>$ of the $\Phi$ field is changed as
\begin{equation}
<\Phi(g_s(z),\overline{g_s(z)})>=<\Phi(z,\bar z)>+
2f\ln{|c_sz+d_s|^2}\,.
\label{schph}
\end{equation}
In this case the
change of  $<\Phi'>=\partial_z<\Phi>$ under the
transformations (\ref{sch})  is given by
\begin{equation}
<\Phi'(g_s(z))>=(c_sz+d_s)^2[<\Phi'(z)>+2f\partial\ln{(c_sz+d_s)}]\,.
\label{schdr}
\end{equation}
To calculate $<\Phi(z,\bar z)>$ we construct the Green function
$K(z,z')$ as
\begin{equation}
K(z,z')=\sum_{(\Gamma)}\frac{1}{(c_\Gamma z+d_\Gamma)^2
[g_\Gamma(z)-z']}\,.
\label{kex}
\end{equation}
In (\ref{kex}) the summation is performed over all the group
products  $\Gamma=\{z\rightarrow g_\Gamma(z)\}$ of
the base transformations (\ref{sch}),
$\Gamma=I$ being included. The $I$ symbol denotes the identical
transformation and $g_\Gamma(z)= (a_\Gamma
z+b_\Gamma)(c_\Gamma z+d_\Gamma)^{-1}$. The above $K(z,z')$
Green function  satisfies the  conditions that
\begin{equation}
K(g_s(z),z')=Q_s(z)^2K(z,z')\,, \quad
K(z,g_s(z'))=2\pi iJ_s'(z)+K(z,z')
\label{k}
\end{equation}
where $Q_s(z)=(c_sz+d_s)$ and $J_s'(z)$ are 1-forms,
$J_s'(z)=\partial_zJ_s(z)$.

To calculate $<\Phi(z)>$ we start with the following relation
\begin{equation}
<\Phi'(z)>=-\int_{C(z)}K(z,z')<\Phi'(z')>\frac{dz'}{2\pi i}
\label{int}
\end{equation}
where the infinitesimal contour $C(z)$ surrounds the $z$ point
in the positive direction. In addition to eq.(\ref{kex}),
eq.(\ref{int}) employes that $<\Phi(z)>$ is a
holomorphic function of $z$. We presume that $<\Phi(z)>$ has
no singularities inside the fundamental region. In this case
one can reduce the right side of (\ref{int}) to the integrals
over the boundaries $C_s^{(+)}$ and $C_s^{(-)}$ of the
fundamental region $\Omega$, both $C_s^{(+)}$ and $C_s^{(-)}$
being given by (\ref{circ}). Furthermore, the $z\rightarrow
g_s(z)$ replacement allows to reduce the integral along the
$C_s^{(+)}$ contour to the integral
along $C_s^{(-)}$. In this case,
employing eqs. (\ref{schdr}) and (\ref{k}), one can derive from
(\ref{int}) that
\begin{equation}
<\Phi'(z)>=-2f\sum_{s=1}^{n}\int_{C_s^{(-)}}K(z,z')
\frac{c_sdz'}{c_sz+d_s}+
\sum_{s=1}^{n}h_s J'_s(z)
\label{fin}
\end{equation}
where $h_s$ are constants. For  $h_s$ to be arbitrary,
the change of (\ref{fin})
under every kleinian group transformation (\ref{sch}) differs
from  (\ref{schph}) by a constant term $H_s$. The above $h_s$
constants are just determined from the conditions that $H_s=0$
for $s=1,2,\cdots,n$.

Eq.(\ref{fin}) gives the desired $<\Phi'(z)>$ value.
It follows from (\ref{fin}) and from (\ref{k}) that
\begin{equation}
<\Phi'(z)>\rightarrow -\frac{2fn}{z}\quad {\rm at}\quad
z\rightarrow\infty
\label{infty}
\end{equation}
At the same time, every non-singular
1-form decreases at $z\rightarrow\infty$ not slower than
$const\cdot z^{-2}$. As example, the $J'_s(z)$ 1-forms
in (\ref{fin}) possess this property. So  one concludes
that $<\Phi'(z)>$
is singular  at $z\rightarrow\infty$.

Owing to the presence of $<\Phi'(z)>$,
an additional
contribution into the vacuum value $<T_\Phi(z)>$ of the 2d
gravity stress tensor $T_\Phi$ arises. To determine
the above contribution, one can substitute in eq.(\ref{tphi})
the $<\Phi'(z)>$ value instead of $\Phi'$. Because of the discussed
contribution, it is appears that
\begin{equation}
<T_\Phi(z)>\rightarrow
-\frac{2f^2n(n-1)}{z^2}\quad {\rm at}\quad z\rightarrow\infty\,.
\label{tinf}
\end{equation}
Eq.(\ref{tinf}) follows from (\ref{tphi}) and (\ref{infty}).
It is useful to remind that  every
non-singular 2-form decreases at $z\rightarrow\infty$ not slower than
$const\cdot z^{-4}$. So, for $n\geq2$,
$<T_\Phi(z)>$  is singular at
$z\rightarrow\infty$.

One could construct the  $<\Phi(z)>$ value to be non-singular at
$z\rightarrow\infty$. In this case $<\Phi(z)>$ appears to be
singular at finite values of $z$. In any case, for $n\geq2$ the
vacuum value $<T_\Phi(z)>$ of the stress tensor $T_\Phi$
turns out to be singular in
the fundamental region on the complex $z$-plane where the
Riemann surfaces are mapped.   In the next Sec. we show
that the above singularity prevents constructing the multi-loop
partition functions.

\section{Problem of the multi-loop partition functions}

Intuition suggests that a singularity of the vacuum value
of the stress tensor does the theory  to be
self-contradictory. To argue this statement in more rigorous manner,
one can note that in the considered theory \cite{df,dk}
the string amplitudes are expected to be independent of
fixed reference metrics. In this case the genus-$n$  partition
functions for $n\geq2$ satisfy the equations
\cite{d} that are  non other than Ward identities.
In \cite{d} the discussed equations have been obtained in
the critical string theory, but the same consideration
can be performed for the non-critical string \cite{df,dk}.

The  partition functions $Z_n(\{q_N,\overline q_N\})$
depend on $(3n-3)$ complex moduli $q_N$ together with
$\overline q_N$, which are complex conjugated to $ q_N$.
The first of the discussed equations for
$Z_n(\{q_N,\overline q_N\})$ is given by \cite{d}
\begin{equation}
\sum_{N}\chi_N(z)\frac{\partial \ln{Z_n(\{q_N,\overline
q_N\})}}{\partial q_N}=<T(z)>-\sum_{N}\frac{\partial}{\partial
q_n}\chi_N(z)
\label{eq}
\end{equation}
and the second equation is obtained from (\ref{eq}) by the
complex conjugation. The $<T(z)>$ value denotes the vacuum
expectation of the stress
tensor (\ref{tens}) and $\chi_N(z)$ are 2-tensor zero modes.
It is worth-while to note that in eq.(\ref{eq}) the ghost
contribution to $<T(z)>$ is calculated in the special ghost scheme
\cite{d}.  Unlike the well known scheme \cite{fms,mart}, in the
considered scheme \cite{d} the ghost vacuum correlator has no
unphysical poles. Among other things, the above correlator takes
into account those contributions to the partition functions,
which are due to both the moduli volume form and ghost zero
modes. In the critical string theory solution of both eq.(\ref{eq})
and its complex conjugated  fully determine the partition
functions apart only an independent of moduli factor. But
in the  non-critical string theory \cite{df,dk} eq.(\ref{eq})
has no solutions.

Indeed, being non-singular in the fundamental region,  every
$\chi_N(z)$ decreases not
slower than $const\cdot z^{-4}$ at $z\rightarrow\infty$.
So the left side of (\ref{eq}) decreases at least as $z^{-4}$ at
$z\rightarrow\infty$.  At the same time, the right part of (\ref{eq})
decreases only as $z^{-2}$, as it follows from eq.(\ref{tinf}).
So in this case there are no the partition functions,
which are consistent with
the requirement that the multi-loop amplitudes are independent of a
choice of fixed references metrics.  It can be proved that
the above equation (\ref{eq}) appears to be
the modular invariant.  So the absence of the solutions of
eq.(\ref{eq}) is meant that there are no the multi-loop partition
functions, which are consistent with the requirement of the modular
invariance. Only the genus-1 partition function can be constructed
to be modular invariant, as it explained in the Sec.4.

\section{One-loop partition function}
We write down the one-loop amplitudes $A_1$ as
\begin{equation}
A_1=\int Z_1(k,\bar k)<V>d\omega d\bar{\omega}
\label{am1}
\end{equation}
where $Z_1(k,\bar k)$ is the genus-1 partition function and
$<V>$ is the vacuum value of the vertex product integrated over the
Riemann surface. The complex period $\omega$ is given in the terms of
the $k$ multiplier in (\ref{mult}) as
\begin{equation}
\omega=\frac{1}{2\pi
i}\ln{k}\,.
\label{omeg}
\end{equation}

As it follows from (\ref{tinf}), for $n=1$ the singularity
in  $<\Phi>$  does not originate the singularity in   the vacuum
value of the stress tensor. In this case  $<\Phi>$
is given by
\begin{equation}
<\Phi>=-f\ln{\left|\frac{(z-u)(z-v)}{(u-v)^2}\right|^2} +const
\label{phi1}
\end{equation}
where $u$ and $v$ are the coordinates of the fixed points of the
Schottky transformation (\ref{sch}), $f^2$ being given by
(\ref{f}).  Indeed, one verifies that (\ref{phi1}) satisfies
(\ref{schph}).  It is follows from (\ref{phi1}) that
\begin{equation}
<\Phi'>=-f\left[\frac{1}{z-u}+\frac{1}{z-v}\right]\,.
\label{phidr1}
\end{equation}
We write down the vacuum value $<T_\Phi>$ of the $T_\Phi$ stress
tensor as
\begin{equation}
<T_\Phi(z)>=<T_\Phi^{(1)}(z)>+T_\Phi^{(2)}(z)
\label{vtphi}
\end{equation}
where $<T_\Phi^{(1)}(z)>$ is calculated in the terms of the vacuum
correlator of the  $\Phi$ fields in the accordance with the rules
\cite{fms} of the conform theory and
\begin{equation}
T_\Phi^{(2)}(z)=-\frac{1}{2}(<\Phi'(z)>)^2+f<\Phi''(z)>
\label{tphi2}
\end{equation}
It is follows from (\ref{phidr1}) and (\ref{tphi2}) that
\begin{equation}
T_\Phi^{(2)}(z)=\frac{1}{2}f^2\frac{(u-v)^2}{(z-u)^2(z-v)^2}\,.
\label{tphi22}
\end{equation}
One can see from (\ref{tphi22}) that $T_\Phi^{(2)}$ has
no singularities in the fundamental region on the complex $z$ plane,
the $z=\infty$ point being included.

Moreover,
instead of the Schottky parameterization,
one can map every  genus-1 Riemann surfaces into the
parallelogram on the complex $w$ plane. In this case the kleinian
group is given by the linear coordinate transformations
\begin{equation}
w\rightarrow w+1\,\qquad w\rightarrow w+\omega
\label{linkl}
\end{equation}
where $\omega$ is the complex period. The above parameterization
can be obtained from that where the Schottky group are employed
by the mapping
\begin{equation}
w=\frac{1}{2\pi i}\ln{\frac{z-u}{z-v}}
\label{trans}
\end{equation}
with both $u$ and $v$ to be the
coordinates of the fixed points of the transformation  (\ref{sch}).
The first of eqs.(\ref{linkl}) corresponds to $2\pi$-twist around the
$C^{(+)}$ or $C^{(-)}$ circle on the complex $z$ plane, both
$C^{(+)}$ and $C^{(-)}$  being given by (\ref{circ}). The second of
eqs.(\ref{linkl}) corresponds to the mapping (\ref{sch}) on the
$z$ plane.

It is follows from (\ref{phichan}), (\ref{phi1}) and
(\ref{trans}) that, the kleinian group  (\ref{linkl}) being
employed, the $<\Phi>$ value is independent of $w$ and,
therefore, $<\Phi'>=0$.  This result is natural. Indeed, one can
see from (\ref{phichan}) that $<\Phi>$ is scalar under the
kleinian group (\ref{linkl}). Besides, $<\Phi'>$ is the
holomorphic function of $z$. Therefore, $<\Phi>$ appears to be
the zero mode of the Laplacian. So $<\Phi>=const$ and
$<\Phi'>=0$.  In this case the one-loop partition function can
be calculated by the method given in \cite{pol}, as it has been
done in \cite{gtw}.

In the above paper \cite{gtw} the $Z_1(k,\bar k)$  partition function
has been obtained to be
\begin{equation}
Z_1(k,\bar k)=const\frac{|k|^{\frac{1}{12}}|y|^2\tilde{Z}_m}
{Im\omega}
\label{p1}
\end{equation}
where $const$ is independent of $k$ and $\bar k$,
$\omega$ is defined by (\ref{omeg}). The
$y=y(k)$ function is given by
\begin{equation}
y(k)=\prod_{n=1}^{\infty}(1-k^n).
\label{y}
\end{equation}
The $\tilde{Z}_m$ multiplier in (\ref{p1}) is due to the string
fields.  In the considered scheme \cite{df,dk}
\begin{equation}
\tilde{Z}_m=\frac{|k^{\frac{1}{24}}y|^{-2d}}{(Im\omega)^{d/2}}\,,
\label{zm}
\end{equation}
$y$ being defined by (\ref{y}). The factor before $\tilde{Z}_m$
in (\ref{p1}) is due to the ghost and $2d$ gravity fields. It has been
noted in \cite{gtw} that $\tilde{Z}_m$ being defined by (\ref{zm}),
is modular invariant and $Z_1(k,\bar k)$ satisfies the condition
of the modular invariance of the one-loop amplitudes.

The same result is obtained in the Schottky parameterization,
the contribution due to $<\Phi'>$ being taken into account. This
proves that the discussed contribution is really necessary to obtain
the correct result.To prove the above statement, we note
that $Z_1(k,\bar k)$ satisfies
eq.(\ref{eq}) just as the multi-loop partition functions do it.  In
this case, however, the ghost scheme \cite{d} should be
modified to take into account the contribution to $Z_1$ due to
the zero modes of the vector ghosts. It could be achieved by a
suitable choice of the ghost vacuum correlator.
But this problem is not
considered in the present paper.  Instead we note that the ghost
contribution to $Z_1(k,\bar k)$  is the same as in the critical
string theory. And we use the critical string calculations
\cite{d,mart,vec} to determine the contribution discussed.

We write down the desired $Z_1(k,\bar k)$ value as
\begin{equation}
Z_1(k,\bar k)=Z_\Phi(k,\bar k) Z_{gh}(k,\bar k) Z_m(k,\bar k)
\label{z1sch}
\end{equation}
where $Z_\Phi,$ $Z_{gh}$ and $Z_m$ are due to the $2d$ gravity field,
the ghost fields and the string fields, respectively.
In this case both $Z_\Phi(k,\bar k)$ and $Z_{gh}(k,\bar k)$
are determined  by the following equations
\begin{equation}
\chi_k(z)\frac{\partial\ln{Z_m}(k,\bar k)}{\partial
k}=<T_m(z)>\,,
\label{eqm}
\end{equation}
\begin{equation}
\chi_k(z)\frac{\partial\ln{Z_\Phi}(k,\bar k)}{\partial
k}=<T_\Phi^{(1)}(z)> +T_\Phi^{(2)}(z)\,,
\label{eqphi}
\end{equation}
their complex conjugated being added, too.
The $<T_\Phi^{(1)}(z)>$ and $T_\Phi^{(2)}(z)$ values are
given by  (\ref{vtphi}) and (\ref{tphi22}).

It is worth-while to note that the $\chi_k(z)$ tensor zero mode
in (\ref{eqm}) and in (\ref{eqphi}) is normalized by \cite{d}
\begin{equation}
\int_{C^{(-)}}\chi_k(z)\frac{(z-u)(z-v)}{k(u-v)}\frac{dz}{2\pi i}=-1
\label{norm1}
\end{equation}
where $C^{(-)}$ circle is defined by eq.(\ref{circ}).
So eq.(\ref{tphi22}) can be rewritten down as
\begin{equation}
T_\Phi^{(2)}(z)=\frac{f^2}{2k}\chi_k(z)\,.
\label{tphi222}
\end{equation}
In the Schottky parameterization the  $<T_m>$ value is calculated
by the same methods  that have been used \cite{d,mart,vec} in the
critical string theory. As the result, the $Z_m$ value being
calculated from (\ref{eqm}) turns out to be
\begin{equation}
Z_m=\frac{|y|^{-2d}}{(Im\omega)^{d/2}}
\label{zmfin}
\end{equation}
where $y$ is given by (\ref{y}). Furthermore, using
eq.(\ref{zmfin}) together with the results given in
\cite{pol}, we obtain that
\begin{equation}
Z_{gh}=\frac{|y|^2}{(Im\omega)|k|^2}\,.
\label{zghfi}
\end{equation}
Furthermore, $<T_\Phi^{(1)}>$ is calculated by the same method as
$<T_m>$. Moreover, $T_\Phi^{(2)}$ is given by (\ref{tphi222}),
$f^2$ being given by (\ref{f}). In this case one can find from
(\ref{eqphi}) that
\begin{equation}
Z_\Phi=\frac{|k|^{(1-d)/12}|y|^2}{(Im\omega)}\,.
\label{zphif}
\end{equation}
Being compared with (\ref{p1}) and with (\ref{zm}), eqs. (\ref{z1sch})
and (\ref{zmfin})-(\ref{zphif}) show that every from the $Z_m$,
$Z_{gh}$ and $Z_\Phi$ values depends on the parameterization of
the Riemann surfaces to be chosen.  This is because the conform
anomaly presents in every of the  $T_m,$ $T_{gh}$ and $T_\Phi$ values.
So every of the above values itself does not appear to be a tensor.
But the whole $Z_1$ partition function being defined by
(\ref{p1}) and (\ref{zm})  coincides with that calculated from
(\ref{z1sch}) and (\ref{zmfin})-(\ref{zphif}).
And the above the partition function satisfies the condition
that the one-lop amplitudes are due to be modular invariant.
At the same time, the construction of the multi-loop partition
functions in the scheme \cite{df,dk} appears to be impossible.

This work is supported by the Russian Fundamental Research Foundation,
under grant No. 93-02-3147.

\end{document}